\documentclass[english,11pt,notitlepage,a4paper]{article}%
\usepackage[T1]{fontenc}
\usepackage[latin1]{inputenc}
\usepackage{textcomp}
\usepackage{epsfig}
\usepackage{subfigure}
\usepackage{babel}
\usepackage{amsmath}
\usepackage{amsfonts}
\usepackage{amssymb}
\usepackage{graphicx}
\usepackage{accents}%
\providecommand{\U}[1]{\protect\rule{.1in}{.1in}}
\makeatletter
\begin{document}

\title{Unstable Dynamics, Nonequilibrium Phases and Criticality in Networked
Excitable Media}
\author{S. de Franciscis, J.J. Torres, and J. Marro\\\textit{Departmento de Electromagnetismo y Física de la Materia,} and\\Institute \textit{Carlos I}\ for Theoretical and Computational Physics,\\University of Granada, Spain.}
\maketitle
\date{}

\begin{abstract}
Here we numerically study models of excitable media, namely, networks with
occasionally quiet nodes and with connection weights that vary with activity
on a short--time scale. The networks global activity show spontaneous (i.e.,
even in the absence of stimuli) unstable dynamics, nonequilibrium phases
---including one in which the global activity wanders irregularly among
attractors--- and $1/f$ \textit{noise} as the system falls into the most
irregular behavior. A net result is resilience which results in an efficient
search in the model attractors space that can explain the origin of similar
behavior in neural, genetic and ill--condensed matter systems. By extensive
computer simulation we also address a previously conjectured relation between
observed power--law distributions and the possible occurrence of a
\textquotedblleft critical state\textquotedblright\ during functionality of
(e.g.) cortical networks, and describe the precise nature of such criticality
in the model.


\end{abstract}

\section{Introduction}

A network is said to have attractors when it can autonomously change its
pattern of overall activity to converge with time towards one case while being
resilient to perturbations. Following psychological observations \cite{hebb}
and formal work by an engineer \cite{Amari} and a physicist \cite{hopfield},
the concept was popular two decades ago as a mathematical tool to explore the
fundamentals of brain tasks attributed to cooperation between many neurons.
According to the, say, \textit{standard model }\cite{Amit}, patterns of
information, corresponding to sets of values for the nodes activity, are
stored in a way that affects the intensities of the edges, representing
synapses, which induces a heterogeneous distribution of the edge weights. The
global activity may then converge towards one of the given patterns when
starting with a degraded version of it. That is, the system exhibits kind of
resilience, often known as \textit{associative memory} ---a property that
mimics the process of recognizing a childhood friend we have not seen for
dozens of years--- which being common to humans is difficult to be efficiently
emulated with computers. Such a remarkable consequence of cooperation is also
relevant to the understanding of complexity in a variety of systems and to
solve actual optimization problems \cite{Amari,Amit,iold1,iold2,6bis}.

The systems of interest in nature do much more than just choosing one out of a
set of patterns and staying in its neighborhood, however (see
\cite{Rabi,Piro,Dante} and references therein). For example, signals from the
heart and cortical neural activities have been successfully interpreted using
non--linear dynamic theory \cite{10ant}-\cite{14fin}, and the standard model
has been generalized along biologically--motivated lines that endow it with
even more interesting behavior \cite{other}-\cite{oth7}. In particular, it was
shown that one may capture some of the observed shaky mechanisms and
instabilities by taking into account two features that seem to characterize
generically excitable media \cite{amletico}, namely, assuming both rapid
activity--dependent fluctuations of the edge weights and the existence of
nodes that are reluctant to a change of state during a time interval after
operation. It is remarkable that incorporating these simple mechanisms into
the standard model has allowed one to recreate \cite{torhibrid} the transient
dynamics of activity as observed in experiments concerning the locust odor
representation \cite{mazor}.

The nervous system is definitely not the only network that exhibits both
varying edge weights and silent nodes at a basic level of observation and, as
a reflection of this at a higher level, \textit{roaming dynamics}
characterized by a continuous wandering among attractors. This occurs in
ill--condensed matter, for instance, whose emerging properties are determined
by \textquotedblleft microscopic disorder\textquotedblright. In fact, it is
sensible to imagine such a disorder is more involved than assumed in familiar
spin glass models. That is, the effective interactions between ions should
certainly be expected to have short time variations ---associated to ion
diffusion, basic chemical reactions, and other local changes concerning
impurities, misfits, fields, rearrangements and strains, etc.--- which would
in general induce nonequilibrium patterns of activity as, for example,
observed in reaction--diffusion systems \cite{MD,21b}. It is likely that the
behavior of genetic networks during biological evolution is another case of
microscopically--induced roaming dynamics \cite{BarYam,Nelson,albert}.
Furthermore, though to our knowledge the relevance of roaming has not yet been
described for other excitable systems, it is noticeable that variability of
connections and occasional lack of individual activity are features that
typically characterize friendship, social, professional and business contacts
\cite{Sun}, the case of the interrelated metabolic reactions that run the
cell, food webs, and transport and communication networked systems, for instance.

In this paper, we describe in detail model phenomenology bearing relevance to
situations with spontaneously unstable dynamics associated to excitability as
described in the two previous paragraphs. By extensive computer simulations,
we show both first and second order phase transitions, characterize the nature
of different \textit{nonequilibrium phases} \cite{MD} that occur as one
modifies the system parameters, study the details of the network activity
dynamics, and determine the conditions in which long--range correlations and
non--Gaussian noise emerge. This results in a systematic study that adds up to
recent efforts trying to understand the origin of the observed relation
between certain statistical criticality and dynamically critical functionality
in neuroscience \cite{Piro,Dante,48,noise,48bis,49,sebas00,plenz}. Our study
in this paper complements analytical study of the simplest limits of the same
model in Ref.\cite{amletico} and related exploratory numerical studies therein.

\section{Definition of model}

Consider a network in which the consequences of the activity changes of each
node above threshold may be sketched by means of a binary variable:
$\sigma_{i}=\pm1,$ $i=1,\ldots,N.$ This is known to suffice in practice to
investigate main effects of cooperation in different contexts \cite{red}. Each
node receives a signal ---alternatively, endures a local field---
$h_{i}(\mathbf{\sigma})=\sum_{j\neq i}w_{ij}\sigma_{j},$ where $\mathbf{\sigma
=}\left\{  \sigma_{{i}}\right\}  $ stands for the global activity and $w_{ij}$
is the weight of the connection between nodes $i$ and $j.$ In the problems of
interest, one may typically single out $P$ patterns of activity, namely,
$\left\{  {\xi_{i}^{\mu}=\pm1}\right\}  $ with $\mu=1,\ldots P,$ that have
some special relevance. The weights then follow accordingly, e.g., by means of
the superposition rule $w_{ij}=\frac{1}{N}\sum_{\mu=1}^{P}\xi_{i}^{\mu}\xi
_{j}^{\mu}.$ This is one of the simplest conditions that transforms the
special $P$ patterns into attractors of dynamics \cite{hebb,Amit}.

Short--time variability of connections will be introduced by assuming that
their weights are given by $\bar{w}_{ij}=\epsilon_{j}w_{ij},$ where
$\epsilon_{j}$ is a stochastic variable. In order to mimic the cases of
interest, this variable should change very rapidly compared with the network
characteristic time scale. Therefore, we shall assume it can be described by a
stationary distribution. This is taken here as $P_{\text{st}}(\epsilon
_{j}|\mathbf{\sigma})=\zeta(\mathbf{\sigma})\,\delta(\epsilon_{j}%
-\Phi)+[1-\zeta(\mathbf{\sigma})]\,\delta(\epsilon_{j}-1).$ That is, with
probability $\zeta(\mathbf{\sigma}),$ which in general depends on the global
network activity, the weights are changed by a factor $\Phi$ but remain
unchanged otherwise. Depending on the value of $\Phi,$ this may simulate nodes
excitability or potentiation or fatigue of the connections as a function of
the degree of order in the system. The standard model corresponds to $\Phi=1.$
Other choices for $P_{\text{st}}(\epsilon_{j}|\mathbf{\sigma})$ have been
investigated \cite{sam2}, including one in which the weights change depending
on the degree of local order, and it seems that these details do not modify
essentially the system behavior. We shall further assume for simplicity that
the relevant probability in $P_{\text{st}}$ is a sort of order parameter,
namely,%
\begin{equation}
\zeta(\mathbf{\sigma})=\zeta\left(  \mathbf{m}\right)  \equiv\frac{1}%
{1+P/N}\sum_{\mu=1}^{P}\left[  m^{\mu}(\mathbf{\sigma})\right]  ^{2}.
\label{mq}%
\end{equation}
Here, $\mathbf{m=}\left\{  m^{1}(\mathbf{\sigma}),\ldots,m^{P}(\mathbf{\sigma
})\right\}  $ is a vector whose components are the $P$ overlaps of the current
state $\mathbf{\sigma}$ with each of the singularized patterns, namely,
$m^{\mu}(\mathbf{\sigma})=\frac{1}{N}\sum_{i=1}^{N}\sigma_{i}\xi_{i}^{\mu}.$

Time evolution is a consequence of transitions $\sigma_{i}\rightarrow\pm
\sigma_{i}$ that we performed with probability $\frac{1}{2}\left\{
1-\sigma_{i}\,\text{tanh}\left[  h_{i}(\sigma)T^{-1}\right]  \right\}  .$
Here, $T$ is a parameter that measures the degree of stochasticity driving the
evolution ---the so--called network \textit{temperature}. Another main
parameter is the fraction, $\rho,$ of nodes which is updated at each unit of
time ---the Monte Carlo (MC) step (per node). For simplicity, we shall assume
here these nodes chosen at random from the whole set of $N.$ In this way, the
result is a situation in between the limits $\rho\rightarrow0$ (sequential or
\textit{Glauber updating}) and $\rho\rightarrow1$ (parallel or \textit{Little
updating}). The case of intermediate $\rho$ better corresponds to those
situations in which due to excitability or other causes, e.g., power economy,
not all the elements are active at all times.

For simplicity, we shall be concerned only with mutually orthogonal patterns.
This is achieved in practice setting every node in $\xi_{i}^{\mu}$ for all
$\mu$ equal to $+1$ or $-1$ independently with the same probability, so that
$\mathbf{\xi}^{\mu}\cdot\mathbf{\xi}^{\nu}\simeq0$ for any $\mu\neq\nu$ in a
large system. (Assuming specific sets of $P$ correlated patterns, which is of
great practical interest, is beyond the scope of this paper that intentionally
understates this model detail.) Then, under some restrictions which strictly
require also the limit $\rho\rightarrow0$ (see \cite{tormod} for technical
details), the conditions so far stated may be taken into account by assuming
\textit{effective} weights:
\begin{equation}
\bar{w}_{ij}=\left\{  1-\frac{1-\Phi}{2}\left[  \zeta\left(  \mathbf{m}%
\right)  +\zeta\left(  \mathbf{m}^{i}\right)  \right]  \right\}  w_{ij},
\label{peso}%
\end{equation}
where the components of $\mathbf{m}^{i}$ are $m^{\mu}(\mathbf{\sigma}%
)-2\sigma_{{i}}\xi_{i}^{\mu}N^{-1}.$ We shall consider in the following this
simplified version of our model which coincides with the general case for any
$\rho>0$ after averaging $\bar{w}_{ij}=\epsilon_{j}w_{ij}$ over the stationary
noise distribution $P_{\text{st}}(\epsilon_{j}|\mathbf{\sigma)}.$ As a matter
of fact, (\ref{peso}) may formally be viewed as any learning prescription,
$w_{ij},$ which is affected by a multiplicative noise ---with correlations
built due to the dependence on $\mathbf{m}.$ Incidentally, connections that
are roughly of this type were recently shown to induce sort of criticality in
(neural) population dynamics \cite{MCN}.

\section{Phases and diagrams}

A main observation concerns the nature of the phases exhibited as one varies
the noise parameter, $\Phi,$ the fraction of active nodes, $\rho,$ the
temperature $T,$ and the load parameter $\alpha=P/N.$ It turns out convenient
to monitor the time evolution of various order parameters \cite{Amit2,gta93};
in particular,%
\begin{equation}
M=\left\langle \left\vert \overline{m^{\ast}}\right\vert \right\rangle
=\frac{1}{N}\left\langle \left\vert \overline{\sum\nolimits_{i}\xi_{i}^{\ast
}\sigma_{i}}\right\vert \right\rangle , \label{FNobs}%
\end{equation}
where the asterisk is the value of $\mu$ that identifies the pattern having
the largest squared overlap, $\left(  m^{\ast}\right)  ^{2},$ and
$\overline{\left(  \cdots\right)  }$ and $\left\langle \cdots\right\rangle $
stand, respectively, for averages over time and over independent realizations
of the experiment (i.e., changing both the initial conditions and the set of
the special, stored patterns). The set of the other overlaps, $m^{\mu}$ with
$\mu\neq\ast,$ may be characterized by:%
\begin{equation}
R=\frac{1}{1+\alpha}\left\langle \overline{\sum_{\mu\neq\ast}\left(  m^{\mu
}\right)  ^{2}}\right\rangle ,
\end{equation}
where the sum is over all patterns excluding the one in Eq. (\ref{FNobs}). We
also monitored the global activity by means of%
\begin{equation}
Q=\frac{1}{N}\left\langle \sum\nolimits_{i}\overline{\sigma_{i}}%
^{\;2}\right\rangle .
\end{equation}
Our values for $M,$ $R$ and $Q$ in the following involve sufficient averages
of independent values to obtain smooth typical behavior, namely, from 200 to
1000 MCS and 50 to 100 systems for static values, and from 10000 to 50000 MCS
and 10 systems for time--dependent values, unless indicated otherwise.

In the standard case $\Phi=1,$ for uncorrelated patterns, the system shows
three phases \cite{Amit2,Amit}:

\begin{description}
\item[(Ph$_{1}$)] \textit{Memory phase}, in which the system evolves towards
one of the given patterns ---often known as \textit{pure }or\textit{ Mattis
states}. The stationary state corresponds to maximum overlap with the
particular pattern, so that $M$ is large while $R$ is small in the stationary
state, namely, $R\sim\mathcal{O}\left[  (P-1)/(N+P)\right]  .$ One also has
that $Q\simeq1$ near $T=0.$ (This case is illustrated by the two top graphs in
figure 1.)

\item[(Ph$_{2}$)] \textit{Mixture phase}, in which a large system converges to
a mixture of pure states, so that it exhibits some order but not associative
memory. Therefore, one may have several relatively large overlaps, which
induces that $0<M<1$ with a lower bound ---due to finite size--- of order of
$1/\sqrt{N},$ while $0<R<\left(  P-1\right)  /\left(  1+\alpha\right)  $ with
a lower bound of order of $\left(  P-1\right)  /\left(  N+P\right)  .$ Also,
$Q\simeq1$ near $T=0.$

\item[(Ph$_{3}$)] \textit{Disordered phase,} in which the system remains
completely disordered as dominated by thermal noise. Then, all the overlaps
oscillate around zero, so that $M\sim\mathcal{O}(1/\sqrt{N})$ and $R$ is of
order $\left(  P-1\right)  /(N+P),$ and $Q\simeq0$ in the stationary state.
\end{description}

\noindent These cases correspond, respectively, to the familiar ferromagnetic,
spin glass and paramagnetic phases that are well characterized in studies of
equilibrium magnetic models.

\begin{figure}
[tbh]
\begin{center}
\includegraphics[
height=6.6792cm,
width=9.5265cm
]%
{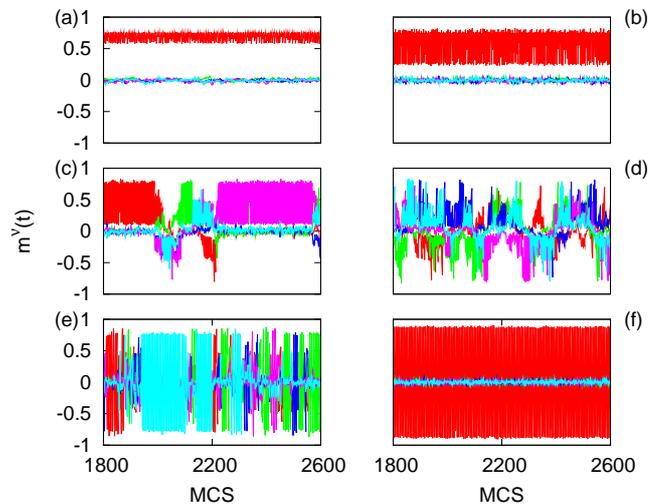}%
\caption{The overlap functions $m^{\nu}(t)$ showing typical different
behaviors for $N=1600$ nodes, $P=5$ patterns, noise parameter $\Phi=-0.5,$
temperature $T=0.01$ and, from top to bottom: associative memory as in
\textbf{Ph}$_{1}$ at (\textit{a}) $\rho=0.10$ (left) and (\textit{b})
$\rho=0.30$ (right); irregular roaming among patterns at (\textit{c})
$\rho=0.375$ (left) and (\textit{d}) $\rho=0.40$ (right) as in \textbf{Ph}%
$_{4}$; eventual jumping between patterns after a set of oscillations between
a pattern and its negative (\textit{antipattern}) as in \textbf{Ph}$_{5}$ at
(\textit{e}) $\rho=0.50$ (left); and pure pattern--antipattern oscillations as
in \textbf{Ph}$_{6}$ at (\textit{f}) $\rho=0.60.$}%
\label{fig1}%
\end{center}
\end{figure}

The behavior of our system is more complex than suggested by this picture,
however. A main novelty for $\Phi\neq1$ is that, as illustrated in figure
\ref{fig1},%
the system exhibits different types of dynamic behavior that cannot be fitted
to the above. That is, one observes that dynamics may eventually destabilize
in such a way that quite irregular jumping ---among attractors as well as from
one pattern to its negative (\textit{antipattern})--- occurs. The observed
behavior suggests one to define the following dynamic scenarios, say,
\textit{nonequilibrium phases }that do not occur in the standard model:

\begin{description}
\item[(Ph$_{4}$)] \textit{Irregular roaming} in which the activity keeps
randomly visiting the basins of attraction corresponding to different
patterns. (This is the case in figures 1(c) and 1(d), i.e., the two middle
graphs in figure 1.)

\item[(Ph$_{5}$)] Irregular roaming as for \textbf{Ph}$_{4}$ but eventually
interrupted at random during some time by oscillations between a pattern and
its antipattern. (This occurs in figure 1(e).)

\item[(Ph$_{6}$)] Pure \textit{pattern--antipattern oscillations}. (As in
figure 1(f).)
\end{description}

\noindent These three genuine nonequilibrium cases correspond to $Q\simeq0$
and $M\simeq0$ (due to orthogonality). Case \textbf{Ph}$_{6}$ also has
$R\simeq0$ (revealing the symmetry of oscillations), while both \textbf{Ph}%
$_{4}$ and \textbf{Ph}$_{5}$ have $R\neq0.$ In order to properly characterize
these dynamic cases, we shall monitor latter the statistics of the itinerant trajectory.

\begin{figure}
[tbh]
\begin{center}
\includegraphics[
height=4.9495cm,
width=7.0387cm
]%
{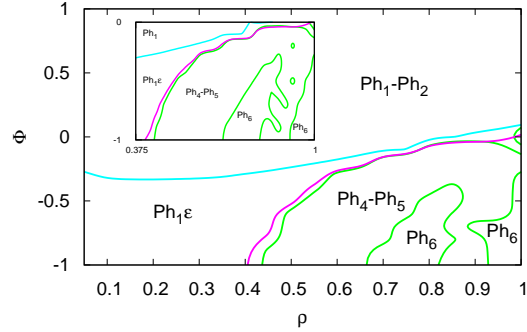}%
\caption{Nonequilibrium phase diagram $\left(  \Phi,\rho\right)  $ at low
temperature. This was obtained for $N=1600,$ $P=5$ and $T=0.1$ from detailed
analysis of all the order parameter functions. The top (blue) line is for
$M=0.8.$ This leaves the equilibrium phases above, where \textbf{Ph}$_{1}$
occurs with probability 0.87 and \textbf{Ph}$_{2}$ otherwise. To the bottom,
the next (violet) line ---leaving also \textbf{Ph}$_{1}\epsilon$ above--- is
for $M=0.5.$ The next (green) lines comprise an inverted--U shaped region in
which $R>0.18.$ The inset shows the roaming region in more detail.}%
\label{fig2}%
\end{center}
\end{figure}

The different behaviors are better observed and interpreted at very low
temperature. As shown in figure \ref{fig2},%
the disordered phase \textbf{Ph}$_{3}$ is not observed at the chosen (low)
temperature, while the ordered, ferromagnetic and spin--glass phases then
occur for any $\Phi$ as far as $\rho$ is not too large. That is, one may have
familiar order as in equilibrium ---practically independently (over a wide
range) of the noise affecting the connections--- as far as only a relatively
small fraction of nodes are simultaneously active \cite{tormod}. However, one
observes small fluctuations or dispersion with time around the mean value $M,$
and that the amplitude of this kind of \textquotedblleft
error\textquotedblright\ increases as one lowers $\Phi$ and increases $\rho.$
This effect, which is evident when one compares the two top panels in figure
\ref{fig1}, led us to indicate a zone \textbf{Ph}$_{1}\epsilon$ around the
region for $\Phi<0$ and $\rho\lesssim0.5.$ It is worth to distinguish this
zone which reveals how the ferromagnetic phase \textbf{Ph}$_{1}$ has
resilience, i.e., a remarkable stability of the attractor to large
fluctuations. These increase monotonously with increasing $\rho$ and/or
decreasing further $\Phi,$ and it finally results in jumping to other
attractors (as in the two middle graphs in figure \ref{fig1}) when more than
one half of the nodes are simultaneously active. This is the origin of the
genuine nonequilibrium cases \textbf{Ph}$_{4}$, \textbf{Ph}$_{5}$ and
\textbf{Ph}$_{6}$. In fact, as shown in figure \ref{fig2}, one observes the
onset of irregular roaming with $R\neq0$ and $M=0$ for $\Phi<0$ and $\rho$
between $0.4$ and $0.6$.

\begin{figure}
[tbh]
\begin{center}
\fbox{\includegraphics[
height=5.6357cm,
width=8.0342cm
]%
{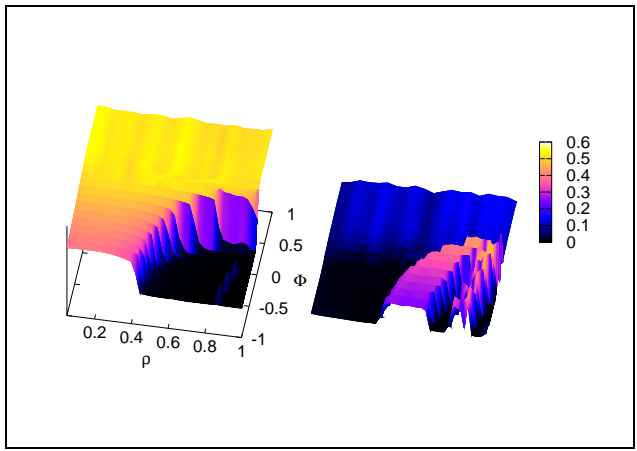}%
}\caption{$M(\Phi,\rho)$ (left) and $R(\Phi,\rho)$ (right; same axes but not
shown for clarity) for $N=1600,$ $P=5,$ and $T=0.1.$ There is coexistence of
\textbf{Ph}$_{1}$ and \textbf{Ph}$_{2}$ for $\Phi>0,$ while the latter phase
does not show up for $\Phi<0$ and memory then occurs but as \textbf{Ph}%
$_{1}\epsilon$ (see the main text) at sufficiently low $\rho.$}%
\label{fig3}%
\end{center}
\end{figure}

\begin{figure}
[tbh]
\begin{center}
\includegraphics[
height=6.3328cm,
width=9.0254cm
]%
{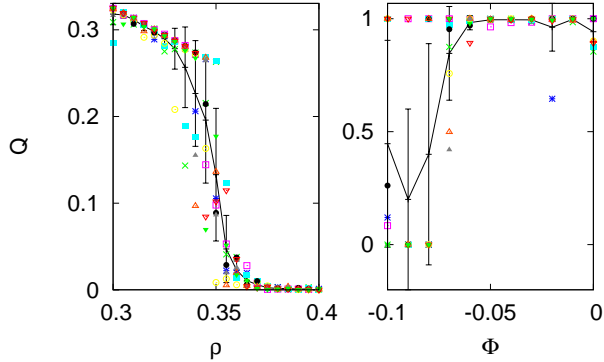}%
\caption{\underline{Left}: Second--order phase transiton between
\textbf{Ph}$_{1}\epsilon$ and \textbf{Ph}$_{4}$ around $\rho\simeq0.37$ when
$\Phi=-0.8.$ \underline{Right}: First--order phase transition between
\textbf{Ph}$_{1}$ and \textbf{Ph}$_{5}$ around $\Phi\simeq-0.1$ when
$\rho=0.8.$ Both plots are for $N=1600,$ $P=5,$ and $T=0.01.$ Note that
different realizations using a different seed produce here different values
corresponding to the different symbols; the mean of all the realizations is
reperesented by a solid curve.}%
\label{fig3bis}%
\end{center}
\end{figure}

\begin{figure}
[tbh]
\begin{center}
\includegraphics[
height=5.6379cm,
width=8.032cm
]%
{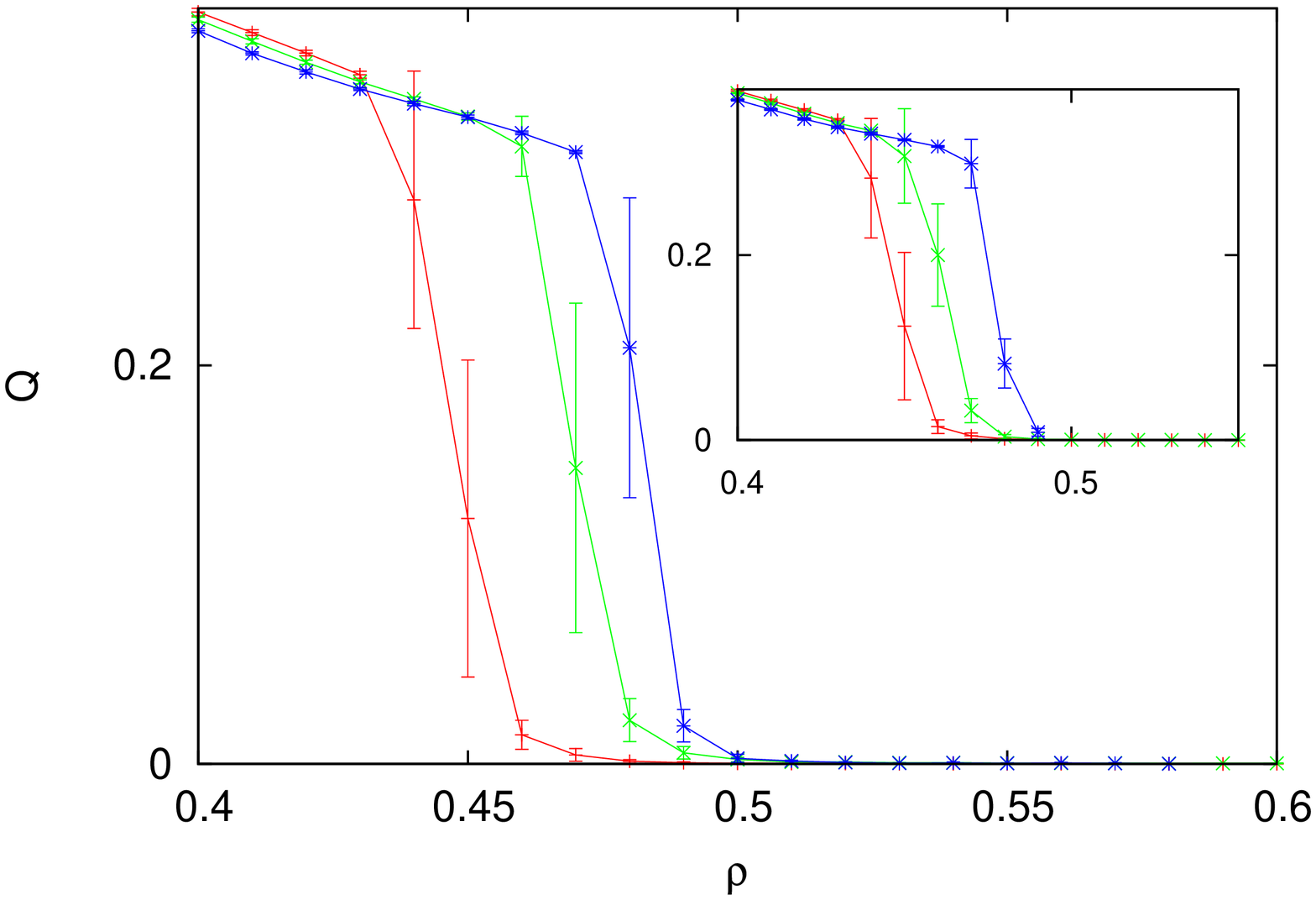}%
\caption{The second--order phase transition on the left of figure
\ref{fig3bis}. For the same system as in this figure, the main graph here
shows data for $P=5$ and $N=1600,$ 3200 and 6400, respectively from left to
right in the middle of the $Q$ value. The inset is for the same values of $N$
but $P=5,$ 10 and 20, respectively, i.e., same value of $\alpha.$ }%
\label{fig3ter}%
\end{center}
\end{figure}

The above picture and figure \ref{fig2} follow from a detailed combined
analysis of functions $M(\Phi,\rho),$ $R(\Phi,\rho)$ and $Q(\Phi,\rho)$ as
illustrated in figure \ref{fig3}. This also shows that two main types of phase
transitions between equilibrium and nonequilibrium phases occur (see figure
\ref{fig3bis}).%
There is a second--order or continuous transition, as one maintains $\Phi<0$
at a constant value, from the memory phase with large \textquotedblleft
error\textquotedblright, i.e., \textbf{Ph}$_{1}\epsilon$, to the irregular
roaming phase\textbf{ Ph}$_{4}$. Then, at least near $T=0,$ one also observes
a first--order or discontinuous transition (figure \ref{fig3bis}),%
as $\rho$ is maintained constant, from the memory phase to the irregular
roaming with pattern--antipattern oscillations, namely, \textbf{Ph}$_{5}$.
Furthermore, it is noticeable here that, as illustrated in figure
\ref{fig3ter},%
the transition region depends on the value of $\alpha=P/N,$ that is, the
critical value of $\rho$ increases somewhat with decreasing $\alpha$ for
finite $N$, and it seems to go to $\rho\simeq0.5$ as $N\rightarrow\infty$ for
finite $\alpha.$

The rare shape of the roaming region in plane $(\Phi,\rho)$ for $P=5,$ which
shows in detail the inset of figure \ref{fig2}, is roughly the same as the one
obtained analytically when $P=1$ for the change of sign of the Lyapunov
exponent in a closely related model (figure 2 in Ref.\cite{amletico}). This
confirms the general observation during our MC experiments of kind of chaos
within the inverted--U region which is delimited in figure \ref{fig2} by the
green lines. That is, one should endow a chaotic character to the roaming
region. That similarity also reinforces the reliability of our measures of
order, and it shows how robust the model here is in relation to the
dynamically irregular behavior. It also follows, in particular, that the model
parameter $P$ is irrelevant to this qualitative behavior, at least as far as
not too many patterns are stored.

\begin{figure}
[tbh]
\begin{center}
\includegraphics[
height=4.9408cm,
width=7.0365cm
]%
{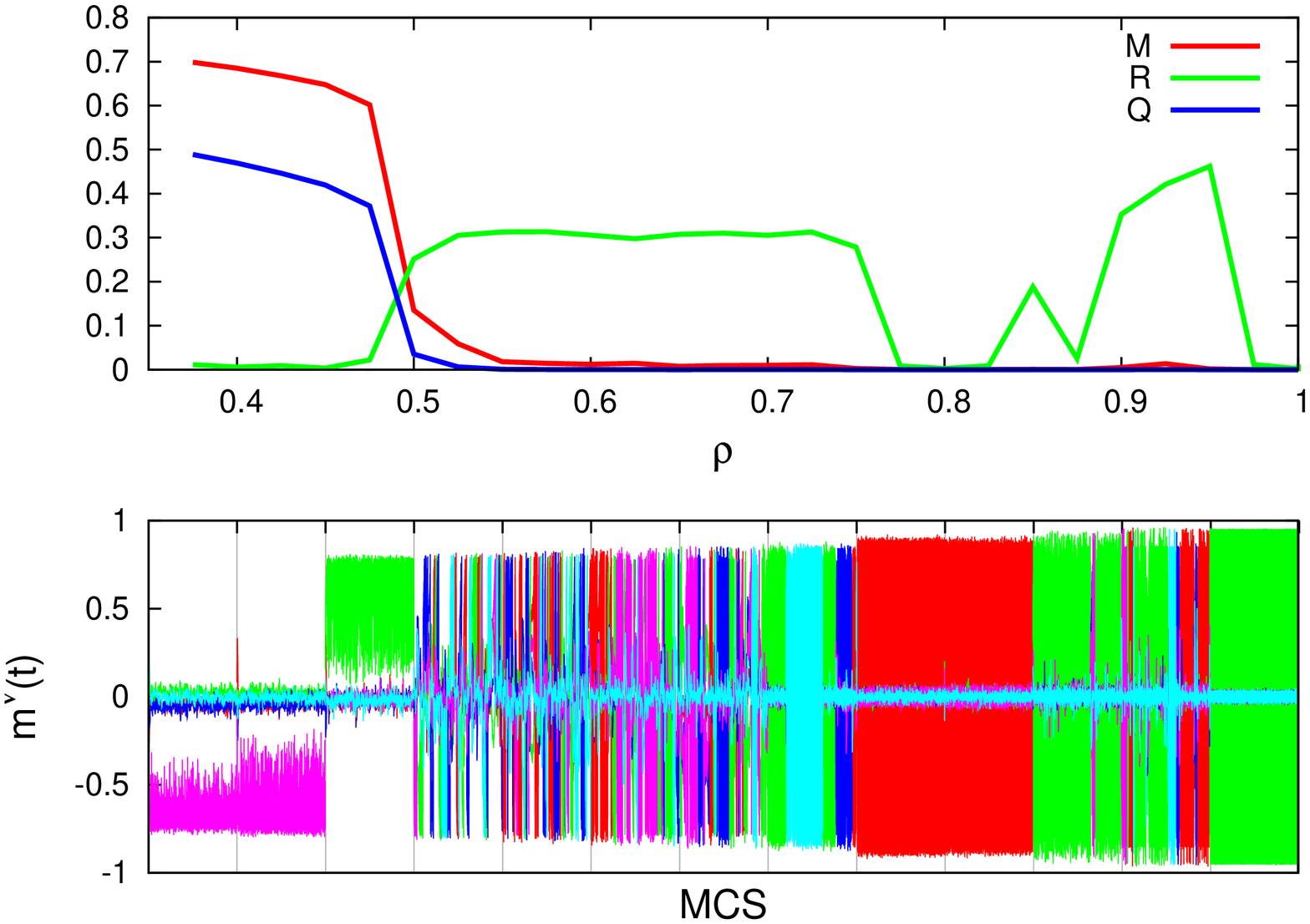}%
\caption{\underline{Upper panel}: Functions $M(\rho)$, $R(\rho)$ and $Q(\rho)$
for $\Phi=-0.7,$ $T=0.1,$ $N=1600$ and $P=5.$ \underline{Bottom panel}: Time
series for the overlap functions $m^{\nu}(t)$ in the same case. The value of
$\rho$ is increased here during time evolution as indicated by the horizontal
axis in the upper panel. Different colours correspond in this graph to
different values of $\nu.$}%
\label{fig4}%
\end{center}
\end{figure}

The \textquotedblleft phases\textquotedblright\ \textbf{Ph}$_{4}$ and
\textbf{Ph}$_{5}$, e.g., cases (d) and (e) in figure \ref{fig1}, cannot be
discriminated on the basis of $M,$ $R$ and $Q$ only. The top panel in figure
\ref{fig4}%
illustrates how these functions change with $\rho$ for fixed $\Phi$ at low
temperature. The bottom panel illustrates the dynamic transition from
irregular roaming in \textbf{Ph}$_{4}$ to the more regular behavior in
\textbf{Ph}$_{5}$ as a consequence of increasing the amplitude of fluctuations
around the attractor as the fraction $\rho$ of active nodes is increased
during time evolution. As indicated in figure \ref{fig2}, the separation
between the memory phase \textbf{Ph}$_{1}$ or \textbf{Ph}$_{1}\epsilon$ and
the nonequilibrium cases is clear cut, while again it results more difficult
to discriminate numerically the region \textbf{Ph}$_{6}$ of pure
pattern--antipattern oscillations (where $M=R=0)$ out of the \textbf{Ph}$_{4}%
$--\textbf{Ph}$_{5}$ chaotic region (where $M=Q=0$ with $R\neq0).$ In any
case, however, our finding concerning this agrees with the analytical result
in a related case \cite{amletico}.

\section{The onset of irregularity}

\begin{figure}
[tbh]
\begin{center}
\includegraphics[
height=5.9843cm,
width=8.5309cm
]%
{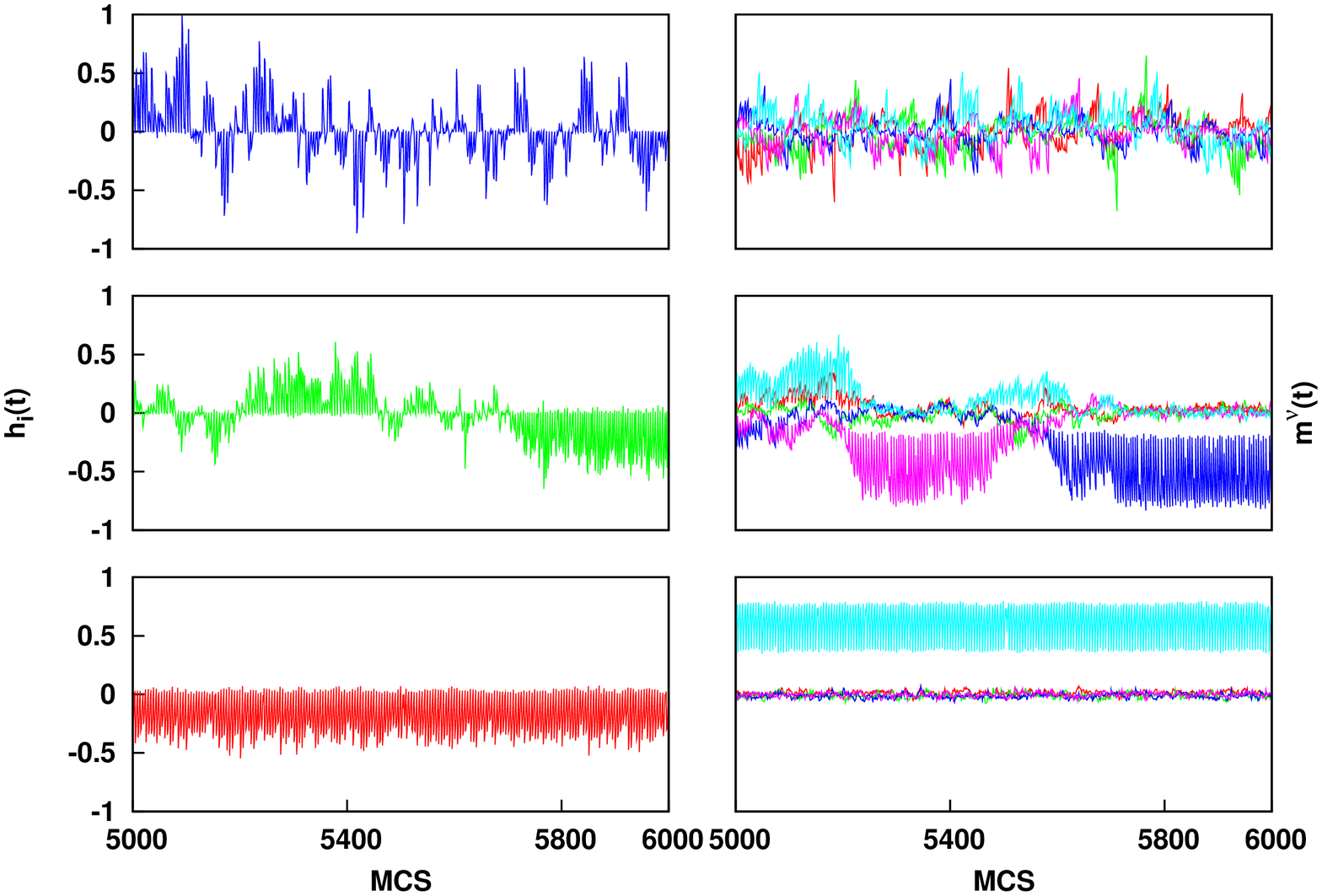}%
\caption{The local signal or field $h(t)$ on a typical neuron (left panels)
and five overlaps $m^{\nu}\left(  t\right)  $ (right panels) indicated with
different colours for a system with $N=1600,$ $P=20,$ $\Phi=-0.80,$ $T=0.01$
and, respectively from top to bottom, $\rho=0.225,$ $0.325$ (near the
transition point), and $0.425$.}%
\label{fig7}%
\end{center}
\end{figure}

The above shows that the most intriguing behavior is when the system activity
becomes irregular, e.g., as one crosses the second--order transition from the
memory phase region to the nonequilibrium behavior ---either at \textbf{Ph}%
$_{4}$ with irregular roaming among attractors or at \textbf{Ph}$_{5}$ where
this may be randomly interrupted by series of pattern-antipattern
oscillations. Figure \ref{fig7}%
illustrates an aspect of this transition. In addition to the time evolution of
some of the overlaps (right panels), which indicates where the activity is at
each moment, this shows (left panels) the signal $h_{i}(t)$ that can sometimes
be monitored in experiments. As a matter of fact, this may be compared, for
instance, with electrical signals measured in single neurons ---as well as
more delocalized, local fields--- in the cortex and hippocampus of rats
\cite{eeg1}, and with MEG signals and recordings for single neuron action
potentials \cite{eeg2,45bis}.

It thus seems it would be most interesting to characterize more quantitatively
how the model signal transforms while performing the relevant transitions.
That is, when moving from the case of random fluctuations around a constant
value in the memory phase, to the case in which the amplitude of the
fluctuations increases and eventually switches to the negative of the original
value, and finally reaches the case in which the frequency of switching and
all the other variables become fully irregular in \textbf{Ph}$_{4}$ and
\textbf{Ph}$_{5}$. With this aim, we studied in detail the distribution of
times of permanence in an interval around significative values of $h.$ More
specifically, in order to extract the relevant information in the case of
quite different signals such as those in figure \ref{fig7}, it turned out
convenient to compute the distribution of time intervals, say $\Delta\tau,$ in
which the signal continuously stays in any of two ranges either $h\left(
t\right)  >h_{0}$ or $h\left(  t\right)  <-h_{0}.$ The cutoff $h_{0}$ intends
to suppress the smallest fluctuations, which correspond to non--significative
noise; this is achieved here in practice for $h_{0}\in\left[  0.05,0.1\right]
.$ We thus observe, after averaging over the network, time and different
experiments that the interesting behavior requires relatively large systems,
so that it does not occur for, say, $N=400$ and $P=5$ while it already becomes
evident for, e.g., $N=6400$ and $P=40.$ The most interesting fact from this
study is that the exponent $\beta$ in a power--law fit $\Delta\tau^{-\beta}$
monotonously increases with size from $\beta\simeq1$ for $N=800$ and $P=10$ in
a way that might indicate a tendency of $\beta$ to 1.5--2 (though our data
never reached this regime). These facts are illustrated in the following figures.

The left panel in figure \ref{fig6}%
\begin{figure}
[tbh]
\begin{center}
\includegraphics[
height=5.6379cm,
width=8.032cm
]%
{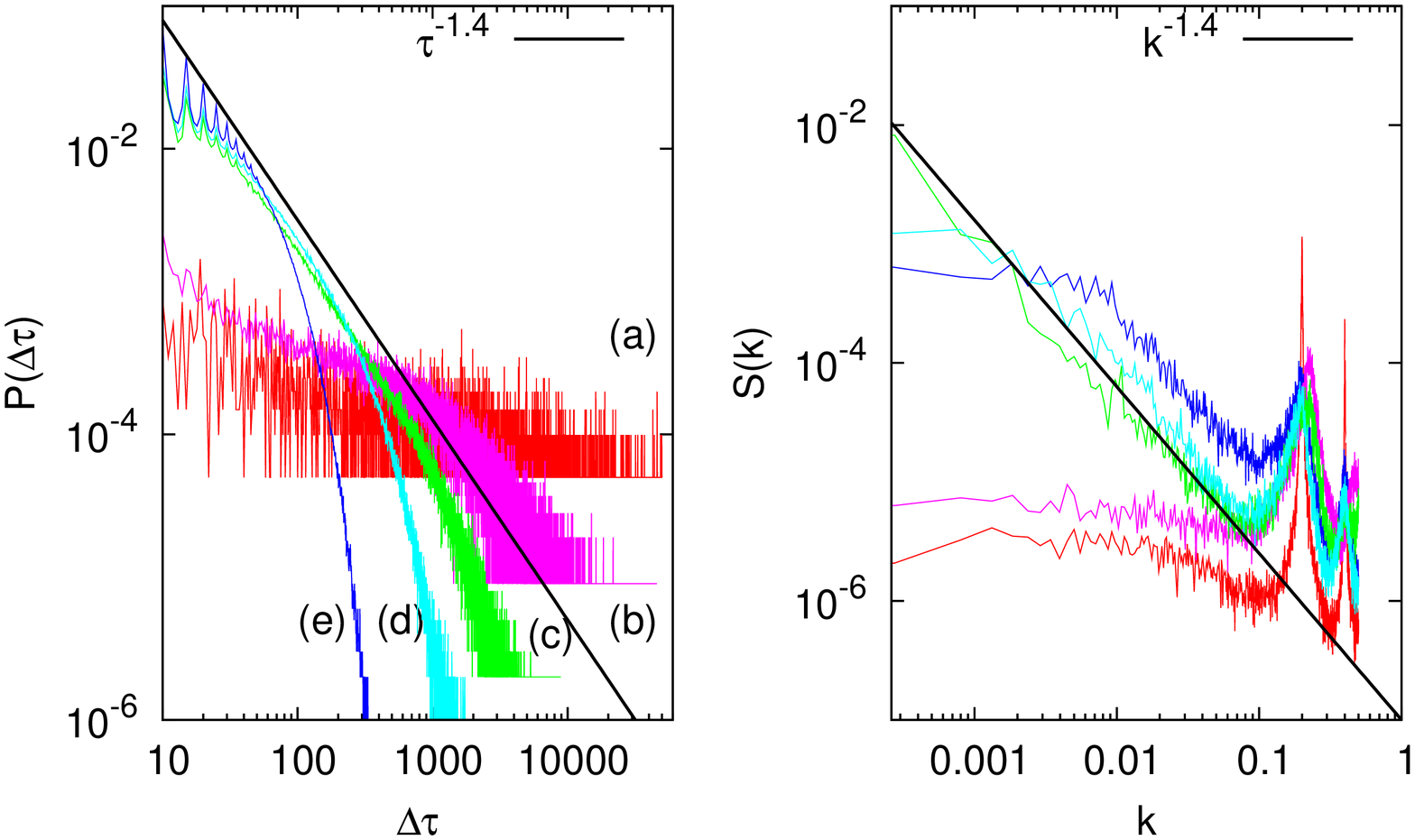}%
\caption{Logarithmic plots. \underline{Left:} Distribution of time intervals
in which the signal continuously stays in any of the two ranges either
$h\left(  t\right)  >h_{0}$ or $h\left(  t\right)  <-h_{0},$ with $h_{0}=0.1,$
when $N=1600,$ $P=20,$ $\Phi=-0.8$ and $T=0.01,$ for the sub--critical cases
$\rho=0.225$ (a) ---a practically horizontal signal in the \textbf{Ph}$_{1}$
phase--- and $0.3$ (b), the super--critical cases $\rho=0.35$ (d) and $0.425$
(e) ---an exponential behavior in the \textbf{Ph}$_{4}$ phase---, and the
near--critical case $\rho=0.325$ (c). The latter, near--critical case
approximately follows the dotted line $\Delta\tau^{-\beta}$ with $\beta=1.4$
for a large time interval. Each case corresponds to an average over 50 neurons
and 20 independent systems running for 10$^{5}$ MCS. \underline{Right}: Power
spectra of $h\left(  t\right)  $ for the same cases as in the left pannel
using runs with 4$\times10^{5}$ MCS. The power--law is illustrated with a
dotted line.}%
\label{fig6}%
\end{center}
\end{figure}
shows a changeover from a general exponential behavior to a power--law
behavior near the interesting second--order phase transition. Analysis of the
Fourier spectra reveals a similar situation, i.e., changeover from exponential
to power--law behavior, concerning both the signal $h(t)$ (right pannel in
figure \ref{fig6}) and the overlap function $m(t).$ Figure \ref{fig6} is a
definite evidence for statistical criticality as one approaches the relevant
transition. On the other hand, figure \ref{fig7a}%
\begin{figure}
[tbh]
\begin{center}
\includegraphics[
height=4.9408cm,
width=7.0365cm
]%
{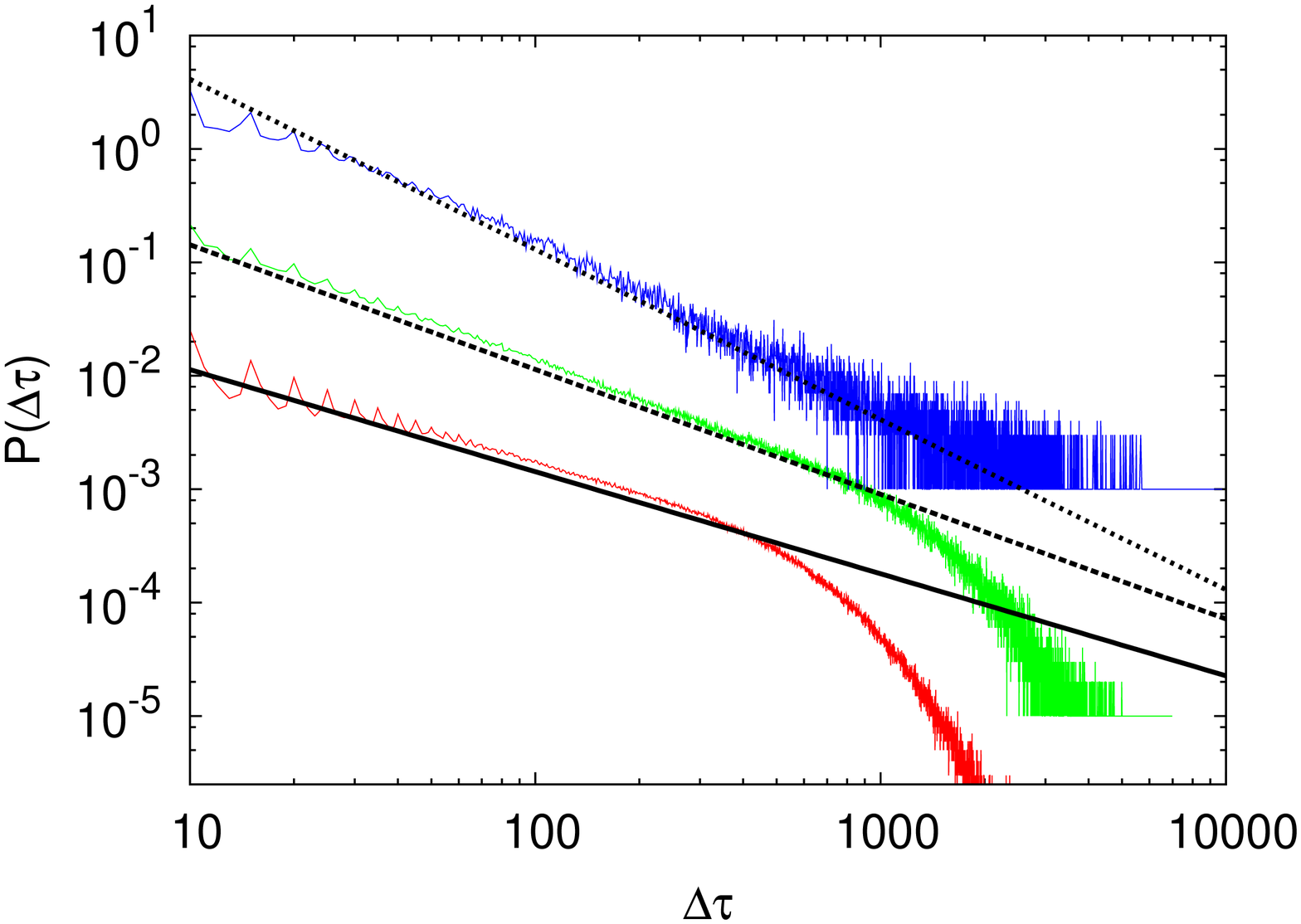}%
\caption{The same as in fig.\ref{fig6} to show the effect of varying the size
$N$ at fixed $\alpha=P/N=0.003125$ and $\rho=0.375.$ From bottom to top, the
data ---corresponding to an average over 50 neurons and 10 independent
systems--- are for $N=1600$ and 3.5$\times10^{6}$ MCS (red), $N=3200$ and
6$\times10^{5}$ MCS (green), and $N=6400$ and 8$\times10^{4}$ MCS (blue),
respectively. (For clarity purposes, there is a vertical translation of the
data points, and it was set $h_{0}=0.05$ here.) Both the exponent $\beta$ in
$\Delta\tau^{-\beta}$ as well as the cutoff at which this power law fails
clearly increase as $N$ is increased.}%
\label{fig7a}%
\end{center}
\end{figure}
shows how the system activity close to the transition between the memory
equilibrium phase\textbf{ Ph}$_{1}$ and the irregular behavior in
\textbf{Ph}$_{4}$ tends to follow the power law distribution over a larger
range as one increases the size $N$ for fixed $P,$ which decreases $\alpha$.
However, we observed (not shown) that $\beta$ does not depend on $N,$ namely,
the same value $\beta=1.4$ is obtained when $P=20$ for $N=1600,$ 3200 and 6400.

\section{Final discussion}

Chemical reactions diffusing on a surface, forest fires with constant ignition
of trees,
parts of the nervous system vigorously reacting to weak stimuli, and the heart
enduring tachycardia are paradigms of \textit{excitable systems} ---out of
many cases in mathematics, physics, chemistry and biology; see
\cite{kaplan,izhi}, for instance. Despite obvious differences, these systems
share some characteristics. They comprise spatially distributed
\textquotedblleft excitable\textquotedblright\ units connected to each other
and cooperating to allow for the propagation of signals without being
gradually damped by
friction. The simplest realization of the relevant excitability consists in
assuming that each element has a threshold and a refractory time between
consecutive responses. In order to deal with a setting which is both realistic
and mathematically convenient, one may suppose the system is networked with
occasionally quiet nodes and connection weights that vary with activity on
short--time scales. As a matter of fact, experimental observations reveal rest
states stable against small perturbations, which correspond to the silent
nodes here, and rapid varying strength of connections, either facilitating or
impeding transmission, which temporarily affect thresholds and may also induce
time lags during response. Furthermore, it is known that such nonequilibrium
setting induces dynamic instabilities and attractors \cite{amletico,sam2}. On
the other hand, we believe it is likely that this modelling of excitable media
may in fact be related to the one by means of partial differential equations
such as when the simple FitzHugh--Nagumo model \cite{nagumo} is used to
represent each unit.

With this motivation, we have studied excitable media by extensive computer
simulations of a discrete time model with an updated rule which generalizes
the Hopfield--like standard case. The resulting phenomenology as described
here is expected to describe the basic behavior in a number of apparently
diverse man--made and natural excitable systems. In particular, we explicitly
show how the model exhibits in the absence of stimuli highly unstable dynamics
when a sufficiently large fraction $\rho$ of nodes are synchronized and for
certain values of a noise parameter $\Phi$ that controls the noise within the
connections strength. We also illustrate how these instabilities induce the
occurrence of novel, first-- and second--order nonequilibrium phases. One of
these happens to be most interesting as it describes the global activity
wandering irregularly among a number of attractors, details strongly depending
on the values of $\rho$ and $\Phi.$ In particular, one may tune an efficient
search in the model attractors space which is sensible to assume it may be at
the origin of phenomenology previously described for neural, genetic and
ill--condensed matter systems. There is also definite evidence of
non--Gaussian, $1/f$ \textit{noise} when the system is tuned into this
irregular behavior, which may explain recent experimental observations of
criticality and power--law distributions in cortical networks.

Finally, we remark how the mechanism behind the irregular jumping from one
pattern to the other is well understood in the model. That is, the relevant
instabilities are to be directly associated to the effective local fields that
one may write as%

\begin{equation}
h_{i}^{\text{eff}}\approx\left[  1-(1-\Phi)\zeta(m)\right]  \sum_{j\neq
i}\omega_{ij}\sigma_{j} \label{eq6}%
\end{equation}
for large $N,$ i.e., neglecting terms of order $N^{-1}.$ After some
manipulation, one may write this more explicitly as%
\begin{equation}
h_{i}^{\text{eff}}=h_{i}^{\text{Hebb}}-\eta\sum_{\mu}\xi_{i}^{\mu}(m^{\mu
})^{3}-\eta\sum_{(\mu\neq\nu)}\xi_{i}^{\mu}m^{\mu}(m^{\nu})^{2}. \label{eq:7}%
\end{equation}
Here, $h_{i}^{\text{Hebb}}$ stands for the energy per neuron in the standard
model, $\eta=(1-\Phi)/(1+\alpha),$ and the last sum is over all pairs of
different indexes $\mu\;\text{and}\;\nu.$ As discussed above, $h_{i}%
^{\text{Hebb}}$ tends to drive the system activity near the attractor
associated to one of the stored patterns. Together with the second term in Eq.
(\ref{eq:7}), this sums up to $\sum_{\mu}\xi_{i}^{\mu}m^{\mu}[1-\eta(m^{\mu
})^{2}]$ which, depending on the value of $\eta,$ induces instabilities and
irregular behavior of the overlaps dynamics similar to those in a cubic map
\cite{cubic}. The third term in (\ref{eq:7}), on the other hand, may be
written as $-\eta\sum_{\nu}m^{\nu}h_{i}^{\nu}$ with $h_{i}^{\nu}=\sum_{\mu
\neq\nu}m^{\mu}\xi_{i}^{\mu}m^{\nu}.$ Given that $\nu$ differs from $\mu$
here, this only includes asymmetric terms $\xi_{i}^{\mu}m^{\nu}$similar to
those that characterize the local fields for asymmetric learning rules,
namely, $\hat{h}_{i}=\sum_{\mu}\xi_{i}^{\mu}m^{\mu+1},$ which are often used
to stored and retrieve ordered sequences of patterns \cite{ordered,gta93}. It
is sensible to assume, therefore, that this term is most efficient in the
present case in inducing transitions among patterns. Unlike for asymmetric
learning \cite{ordered}, however, the destabilization here does not induce any
order nor predictability in the sequence of visited patterns.

This work was supported by the Junta de Andalucía project FQM--01505, and by
the Spanish MICINN--FEDER project FIS2009--08451.


\begin{thebibliography}{99}
\bibitem {45bis}See, however, Ref.\cite{14fin} for an important concern on
temporal and spatial resolution.
\end{thebibliography}
\end{document}